\documentclass[letterpaper, 10 pt, journal, twoside]{IEEEtran}
\usepackage{cite}
\usepackage{textcomp}
\usepackage{wrapfig}

\usepackage[cmex10]{amsmath}
\usepackage{amssymb}
\usepackage{amsfonts}
\usepackage{algpseudocode}
\usepackage{amsthm}
\usepackage{graphicx}
\usepackage{epsfig}
\usepackage{cite}
\usepackage{tensor}
\usepackage{color}
\usepackage{multirow}
\usepackage[ruled,vlined]{algorithm2e}
\usepackage{array}
\usepackage{xcolor}
\usepackage{hyperref}
\hypersetup{
    colorlinks=true,
    linkcolor=blue,
    filecolor=magenta,      
    urlcolor=blue,
    }

\graphicspath{{figures/}}
\DeclareGraphicsExtensions{.eps}

\theoremstyle{definition}

\newcolumntype{P}[1]{>{\centering\arraybackslash}p{#1}}
\newcolumntype{M}[1]{>{\centering\arraybackslash}m{#1}}

\renewcommand{\baselinestretch}{0.988}   

\title{\bf \LARGE A Multi-Sensor Interface to Improve the Learning Experience\\in Arc Welding Training Tasks
}

\author{Hoi-Yin Lee, \IEEEmembership{Student Member, IEEE}, Peng Zhou, \IEEEmembership{Student Member, IEEE}, Anqing Duan, \IEEEmembership{Member, IEEE},\\Jiangliu Wang, Victor Wu and David Navarro-Alarcon, \IEEEmembership{Senior Member, IEEE}
\thanks{This work has been submitted to the IEEE for possible publication. Copyright may be transferred without notice, after which this version may no longer be accessible \url{https://journals.ieeeauthorcenter.ieee.org/become-an-ieee-journal-author/publishing-ethics/guidelines-and-policies/post-publication-policies/}}
\thanks{This work is supported by the Research Grants Council (RGC) under grant 15212721. \textit{Corresponding author: D. Navarro-Alarcon.}}
\thanks{H.-Y. Lee, P. Zhou, A. Duan, V. Wu and D. Navarro-Alarcon are with The Hong Kong Polytechnic University, Department of Mechanical Engineering, Hung Hom, KLN, Hong Kong. (contact e-mail: dna@ieee.org)}
\thanks{J. Wang is with the CUHK T Stone Robotics Institute, The Chinese University of Hong Kong, Shatin, NT, Hong Kong.}
}

\begin{document}

\markboth{}{Lee 
\MakeLowercase{\textit{et al.}}: A Multi-Sensor Interface to Improve the Learning Experience in Arc Welding Training Tasks}
\maketitle

\begin{abstract}
This paper presents the development of a multi-sensor user interface to facilitate the instruction of arc welding tasks. 
Traditional methods to acquire hand-eye coordination skills are typically conducted through one-to-one instruction where trainees must wear protective helmets and conduct several tests. 
This approach is inefficient as the harmful light emitted from the electric arc impedes the close monitoring of the process; Practitioners can only observe a small bright spot. 
To tackle these problems, recent training approaches have leveraged virtual reality to safely simulate the process and visualize the geometry of the workpieces. 
However, the synthetic nature of these types of simulation platforms reduces their effectiveness as they fail to comprise actual welding interactions with the environment, which hinders the trainees' learning process. 
To provide users with a real welding experience, we have developed a new multi-sensor extended reality platform for arc welding training. 
Our system is composed of: (1) An HDR camera, monitoring the real welding spot in real-time; (2) A depth sensor, capturing the 3D geometry of the scene; and (3) A head-mounted VR display, visualizing the process safely. 
Our innovative platform provides users with a ``bot trainer'', virtual cues of the seam geometry, automatic spot tracking, and performance scores. 
To validate the platform's feasibility, we conduct extensive experiments with several welding training tasks. 
We show that compared with the traditional training practice and recent virtual reality approaches, our automated multi-sensor method achieves better performances in terms of accuracy, learning curve, and effectiveness.
\end{abstract}

\maketitle

\section{Introduction}
\label{sec:introduction}
\IEEEPARstart{A}{rc} welding is one of the most common material fusing methods in modern manufacturing \cite{ghosh2017pulse}. 
In its most basic form, it uses a controllable electric current to melt a joining metal that, once cooled, binds two metallic parts together \cite{khan2007welding}.
Due to its strong and enduring joining properties, 
welding is used across numerous economically important fields, such as automotive and aerospace industries, shipbuilding, steel construction, oil and gas pipelines, to name a few instances \cite{latella2021analysis}.
Despite its widespread use, teaching and learning a proper welding technique has historically presented many challenges to both instructors and trainees \cite{liu2014tutorial}.

\begin{figure}[t]
    \centering
    \includegraphics[width=1\linewidth]{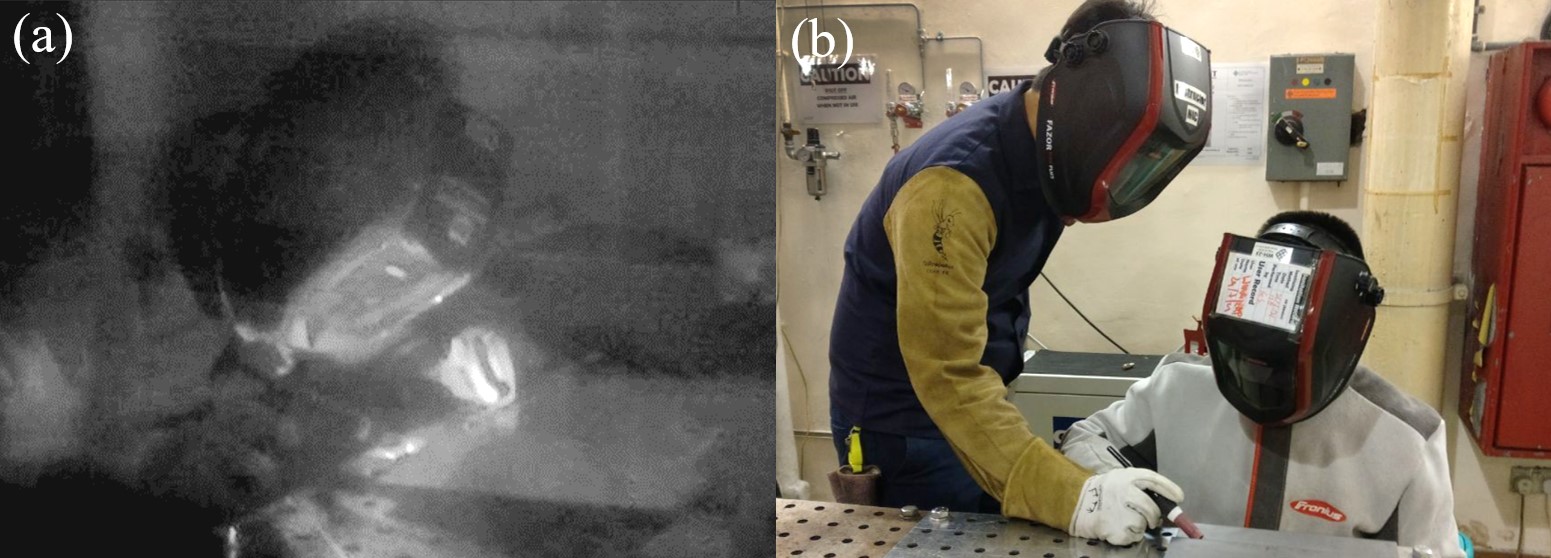}
    \caption{(a) Arc welding; (b) Illustration of traditional training approaches, where the trainer needs to demonstrate the welding skill one by one. 
    }
    \label{fig:teach}
\end{figure}

Traditional approaches to acquire welding skills are typically conducted as follows: The instructor first explains the working principle of the process and demonstrates it with sample welds \cite{asplund2020lessons}; Trainees proceed to conduct hands-on welding tasks under the guidance from the instructor, see Fig.~\ref{fig:teach}. 
The main limitation of this approach comes from the inability to clearly observe the workpiece geometry and the process through the protective helmet, which must be worn at all times \cite{weman2011welding,antonini2003health,althouse2004modern} (beginners generally struggle to obtain a clear spatial notion of what is occurring on the other side of the helmet). 
The near-dark experience makes it difficult to learn how the different configurations affect the quality of a weld; It is hard for trainees to understand a process they can barely observe. 
Furthermore, as there is a limited number of instructors that generally participate in a session, trainees cannot receive real-time support during their practice (advice is received only after the task has been completed and the helmet removed). 
All these factors complicate the instruction and skill acquisition of arc welding tasks.

To address the limitations of traditional training approaches, researchers have developed various didactic platforms based on virtual/mixed reality (VR/MR) systems
\cite{chung2020research,huang2020research,isham2020mobile,yang2010virtual,kobayashi2003skill,vergel2020comparative}. 
These platforms are typically designed to help beginners to get familiar with the procedure and its torch movements through the use of virtual environments that simulate the \emph{high-energy} welding process. 
As no actual interaction with the environment occurs, these simulation-based systems can only provide a synthetic perceptual experience to the users, which may hinder the psychological adaptation that practitioners acquire by conducting real tasks in the field (e.g., fear management, thermal and noise sensations, etc. \cite{agrawal2020augmented}).

\begin{figure}[t]
    \centering
    \includegraphics[width=1\linewidth]{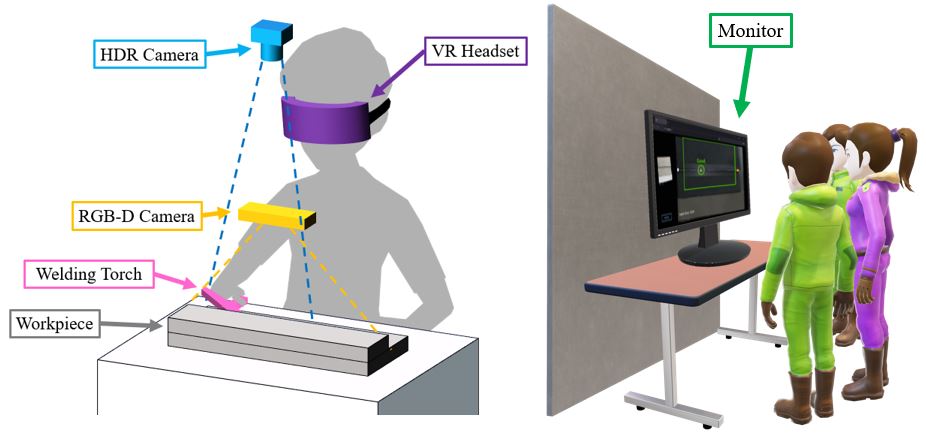}
    \caption{Conceptual representation of the proposed  training system.}
    \label{fig:setup_draw}
\end{figure}

To address the above mentioned issues, in this paper we propose an automated extended reality (XR) training assistant platform that provides the user with a multimodal experience. 
The proposed system consists of a high dynamic range (HDR) camera \cite{wu2020exposure, purohit2021enhancing} to monitor the welding spot in real-time, an RGB-D sensor \cite{yu2018robust, huo2021sensor} to capture the 3D geometry of the scene, and a VR headset to safely visualize the process; Images from these sensors \cite{yang2018novel} are registered into a common frame.
The system uses 3D vision to detect the welding region and seam, and to automatically generate the desired welding path \cite{tian2020automatic}.
These visual cues from the virtual assistant provide the user with valuable extended reality guidance on the weld motion in real-time.
The proposed automated ``bot trainer'' provides the user with a scoring system that quantifies the performance of a task upon its completion. 
As visual information is displayed on a screen in XR mode, the process can be simultaneously observed by the trainee, instructor, and other participants, see Fig. \ref{fig:setup_draw}.

The goal of this XR system is to help trainees to learn the proper torch movements and gain confidence with real welding tasks while reducing the required human instruction to a minimum.
We conducted a series of experiments to evaluate our proposed approach in terms of usability, performance, learning curve, and teaching/learning effectiveness.

Compared with existing approaches, the developed automated training system has the following original features:
\begin{itemize}
  \item It detects and overlays a virtual welding path over the workpiece for trainees to follow.
  \item It provides the user with instantaneous motion recommendations to improve the task performance.
  \item It quantifies and visualizes the performance of the conducted welding task based on sensory feedback.
  \item It enables to display the process in real-time to the user and other participants in the training session.
\end{itemize}

The rest of this paper is organized as follows: Sec. \ref{sec:methods} introduces the architecture of the system. Sec. \ref{sec:results} presents the experiments. Sec. \ref{sec:conclusion} gives the final conclusion.

\section{Methods}\label{sec:methods}
\subsection{System Overview}\label{sec:system_overview}
\begin{figure}[t]
    \centering
    \includegraphics[width=1\linewidth]{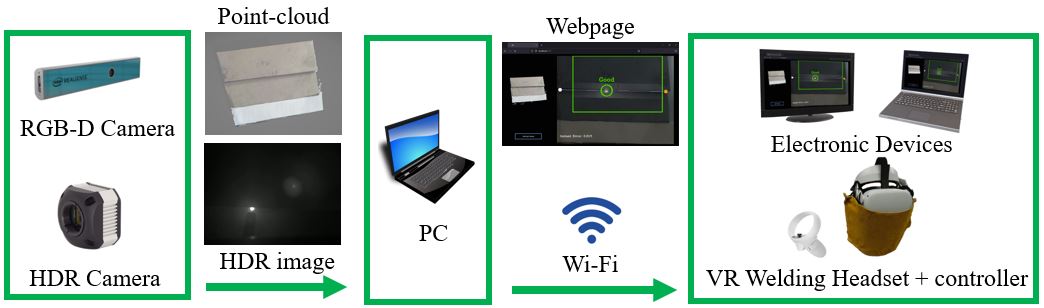}
    \caption{Workflow between the different components in the system. }
    \label{fig:dataflow}
\end{figure}

The data flow among the different components of the XR bot trainer is shown in Fig. \ref{fig:dataflow}. 
It includes an RGB-D sensor, an HDR camera, a VR headset with a controller, a computer, and a welding torch. 
Cameras are cross-calibrated and placed at around 10-degrees from the surface normal and facing towards the welding region. 
Multimodal visual feedback is sent to the computer for processing, and then fed into the VR headset to provide the user with 3D models of the workpiece and a video stream of the welding process.

The workflow depicted in Fig. \ref{fig:dataflow} describes the following steps: Before the process begins, (1) the RGB-D sensor captures a depth image of the welding area to produce a point cloud of the workpiece; (2) A depth-based localization algorithm segments the area of interest and computes the welding seam/path \cite{zhou2021path,Peng2020APC}. 
During welding, (3) the HDR camera captures greyscale images of the process and traces the center of the electric arc in real-time, where (4) all visual information is registered into a 2D image and streamed to a webpage; (5) The VR headset access the 2D live streaming video and the 3D model via Wi-Fi.

The electrode of the welding torch generates a strong electric arc (whose emitted light is harmful to the human eyes) when approaching an electrified workpiece. 
The position of this bright spot is tracked by the system, and its trace is visualized to the user to provide valuable visual cues. 
Upon completion of the task, an evaluation metric is calculated to assess the overall performance; The welding trajectory, average error, and score are displayed on the interface to quantify the performance of the user.

\subsection{Seam Localization}\label{sec:seam_localization}
\begin{figure}[t]
    \centering
    \includegraphics[width=1\linewidth]{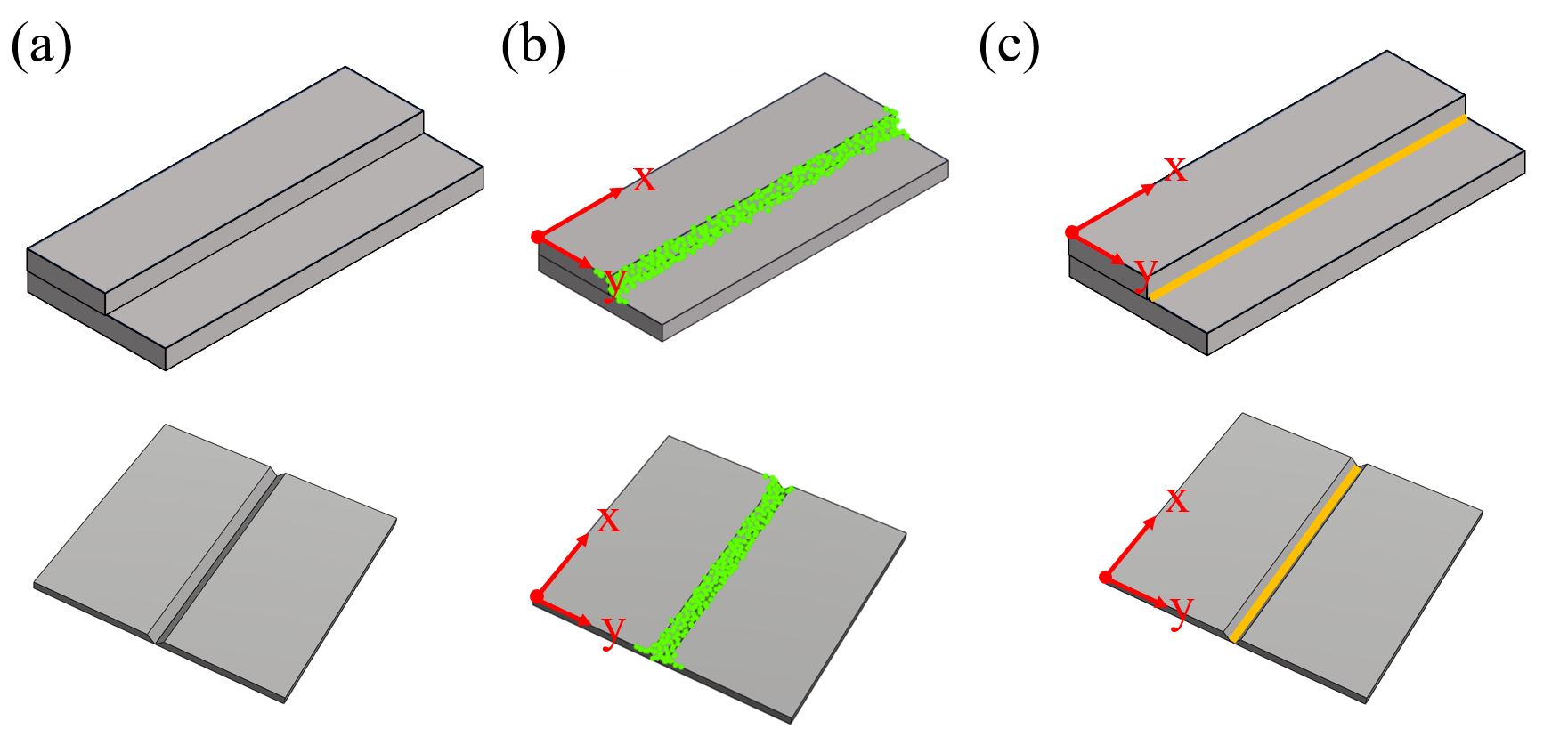}
    \caption{(a) Two different types of workpieces: Fillet welding and Butt welding workpieces; (b) The groove detection algorithm is applied to locate the possible welding area. Groove points \(G\) are displayed in green color; (c) Welding seams are indicated in orange.}
    \label{fig:pt_cloud_draw}
\end{figure}

\begin{figure}[t]
    \centering
    \includegraphics[width=1\linewidth]{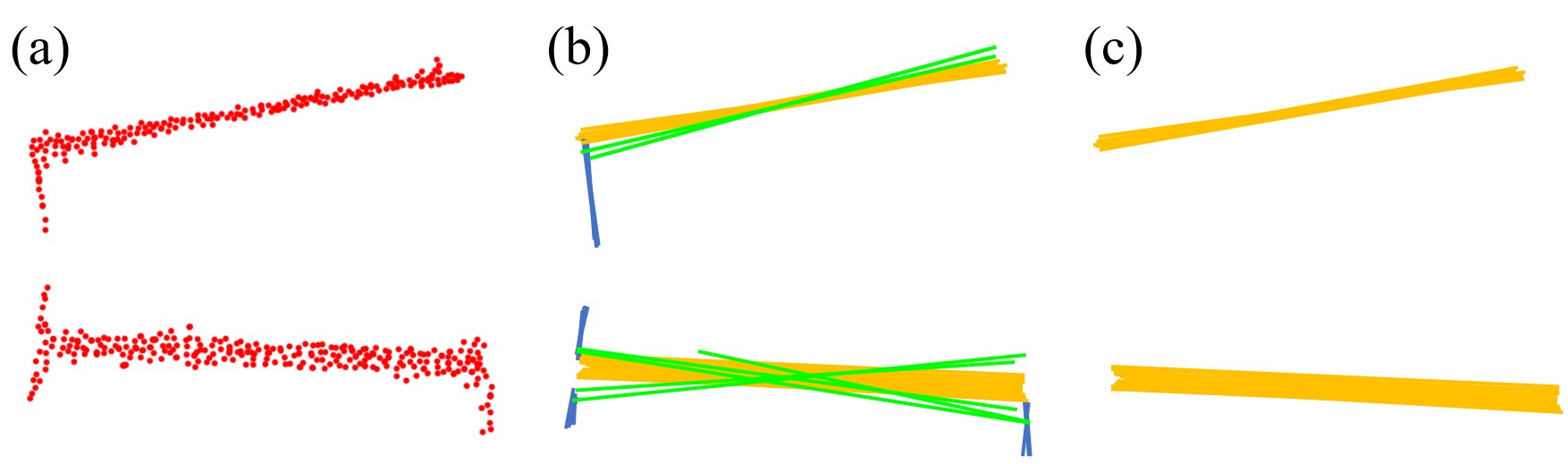}
    \caption{(a) 3D coordinates of the groove  \({G} \) are projected onto a 2D image \(\mathbf {I} \) in red; (b) The computed lines around the seam are the orange and green long lines; Blue lines represent the side edges of the workpiece, which are generally shorter; (c) Lines in blue and green in (b) are removed due to its length and slope.}
    \label{fig:red_dot}
\end{figure}

A groove is a channel between the edges of two metal workpieces \cite{american2010aws}. 
The RGB-D sensor captures an initial color-depth image that covers the whole welding area of the workpiece; This sensor data forms a point cloud \(\mathbf{P} \) of the scene. 
The proposed detection algorithm uses the difference in edge intensity to automatically find the path of the groove, where local neighborhoods in \(\mathbf{P} \) are created to segment the groove's approximated location \cite{zhou2021path,Peng2020APC}. 
In Fig. \ref{fig:pt_cloud_draw}, the groove points \(G = \{g_1,\ldots, g_n\} \) are conceptually presented in green color over the channel. 
Computing the bulk of data from this 3D point cloud results in a long processing time; Several points are involved in finding the seam, which (for beginners) is typically a simple straight line. 
Therefore, the following efficient method for locating the seam is implemented: After the groove has been segmented in the image, the seam is calculated by projecting all points in \({G} \) into a 2D image \(\mathbf{I} \), as represented with red color in Fig. \ref{fig:red_dot}. 
The groove's edges are then calculated from 2D coordinates.

Noisy data from sensor measurements may affect the precision of the seam localization, thus, a kernel convolution noise filter is implemented on \(\mathbf{I} \) to remove noisy/redundant points. 
Canny edge detection \cite{ding2001canny} is then applied to \(\mathbf{I} \) to sharpen the edges of the groove. 
Inspired by the method in \cite{xiong2019robust} (which uses an edge map to extract line segments), we compute a point-line connection edge map to improve the precision of our results. 
Points are interconnected with line segments, as depicted in Fig. \ref{fig:red_dot}.
Euclidean distance \(\lVert{a-b}\rVert\) is then calculated to determine the length of lines with endpoints \(\mathnormal{a} \) and \(\mathnormal{b} \). 
As the welding seam is expected to be longer than the sides of the workpiece, our method removes all short lines. 

To extract the characteristic line $l$ representing the seam, we compute the average slope of $n$ (long) line segments as:
\begin{equation} \label{eq1}
    \overline{s} = \frac{1}{n}\sum_{i=1}^{n} s_i
\end{equation}
for \(\mathnormal{s}_i \) as the slope of the $i$th line segment. 
Lines with a slope outside a range $[\overline{s} - c , \overline{s} + c]$, for \(\mathnormal{c}>0 \) as clearance scalar, are discarded.
This enables us to narrow down the possible seam region to the orange region depicted in Fig. \ref{fig:red_dot}.

A KD-Tree algorithm \cite{bentley1975multidimensional} is introduced to classify the endpoints \(a,b=\mathnormal{E(x,y)} \) of these lines into groups according to their image quadrants. 
The two quadrants that contain most of the endpoints \(\mathnormal{E} \) indicate the orientation of the seam. 
The mean value of \(\mathnormal{E} \) in these two quadrants is used to solve the seam characteristic line \(\mathnormal{l} \). By mapping the depth information in \(\mathbf{P} \) related to the points in \(\mathnormal{l} \), the respective 3D coordinates \(\zeta = \{\zeta_1,\ldots, \zeta_n\} \) can be obtained.
Segments are assigned to \(\mathnormal{l} \) based on the distance between \({\zeta_1} \) and \({\zeta_n} \). 
A point \(Q\) is added to visually indicate in the interface's screen the target to be followed by the user during the task.

\subsection{Electric Arc Localization}\label{sec:hdr}

\begin{figure}[t]
    \centering
    \includegraphics[width=1\linewidth]{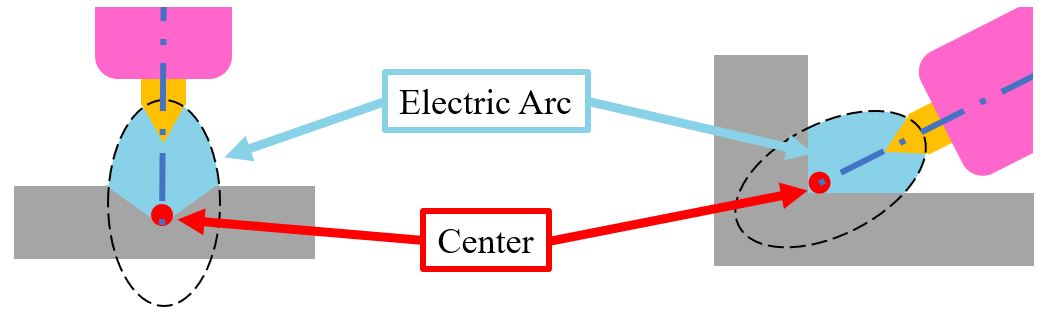}
    \caption{Overview of the electric arc and its center during the process.}
    \label{fig:arc_center}
\end{figure}

Once the welding process starts, a strong light is generated by the electric arc \cite{weman2011welding}, which is perceived through a helmet as a small bright spot; An effective welding task requires to move this spot in the correct direction, i.e., along the seam.
By computing the arc's center (shown in Fig. \ref{fig:arc_center}), the intersection point between the center of the electrode and the seam can be located by our multi-sensor system.
As the intensity of the HDR image is affected by the strong light, thus, only the region around the seam is processed and displayed to the user in real-time. 
Areas far from the seam are statically displayed based on the scene's initial observation (i.e., before the welding process began). 
The region around the electric arc has high-intensity levels in the captured images, thus, this area of interest is extracted with a binarization approach that generates a feature image. 
Nevertheless, some light reflections may still appear in the binary map as its computation is susceptible to noises \cite{xiong2019robust}.

\subsubsection{Light Intensity-Based Confidence Map} \label{sec:LIPMap} 
\begin{figure}[t]
    \centering
    \includegraphics[width=1\linewidth]{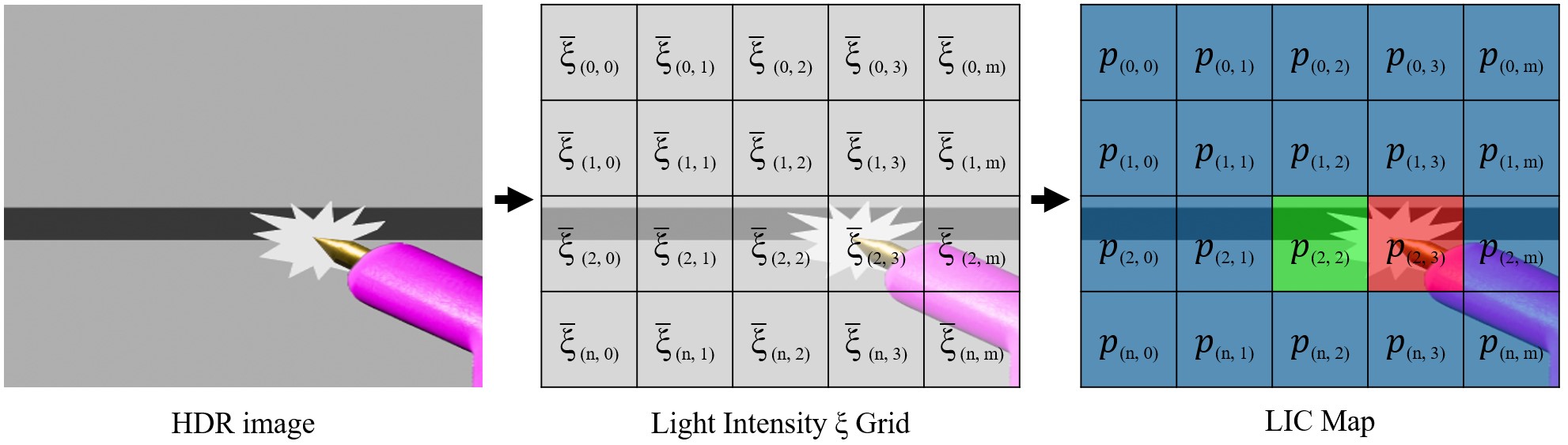}
    \caption{Average light intensity $\bar{\xi}$ is calculated for each grid over the image. An updated LIC ${p[t]}$ is solved for each grid based on the past confidence level $p[t-1]$ and the current light intensity $\bar{\xi}$ value. Three colors are used to visualize the probability of the potential welding spot location. The red color tile indicates it has the highest confidence level; Green indicates high confidence level; Blue indicates low confidence.}
    \label{fig:probMethod}
\end{figure}

To remove flashing noises and minimize the computational time, a confidence map is constructed for each image  \cite{fang2019robust, yasarla2020confidence, idrees2013multi}.
A light intensity-based confidence (LIC) map is applied to the HDR image, where the confidence is determined by the past trajectory and the present observations, see Fig. \ref{fig:probMethod}. 
The frame is first divided into multiple tiles, with an average light intensity $\bar{\xi}$ computed for each of them. 
The light emitted from the welding spot makes the light intensity of the surroundings increase radially from the center of the electric arc. 
As we consider the slow motions of the welding torch by the user, it is reasonable to assume that for a continuous high intensity in a tile for the past few frames, there is an exponentially high likelihood that an equal or greater light intensity value will be obtained in that same tile in the next frame. 
Similarly, for a continuous low intensity in the past few frames, it is unlikely there will be an electric arc shown in the coming frames. 

Our method aims to reduce the probability density distribution of the whole image. 
Only high values are considered to create a high contrast image; 
Thus, an exponential growth/decline is implemented to model the likelihood for the welding spot to be located in a tile of a frame. 
The confidence level of each grid is then calculated with the following function:
\begin{equation} \label{eq_prob}
    {p}[t] = \text{min}\bigg(\frac{(b \cdot \text{norm}(\bar{\xi}))^{p[{t-1}]}}{b}(\sigma + p[{t-1}]), 1 \bigg)
\end{equation}
where $\text{norm}(\bar{\xi}) = {\bar{\xi}}/{255}$ represents the normalized average intensity in each tile; $b>0$ denotes a user-defined parameter to specify the desired intensity range; $\sigma>0$ controls the sensitivity to the current observation, where small values result in delayed responses and large values make it susceptible to fast changes (a good compromise is $\sigma =1$, which is used for the rest of our derivations).
The current probability (at time instance $t$) is denoted by $p[t]$, which describes the confidence of a welding spot to be found in a tile; The method is initialized as $p[t-1] = 0$ for $t = 0$. 
The minimum value in \eqref{eq_prob} ensures that $0\le p[t] \le 1$ is satisfied.

Different torch poses may affect the light intensity values, making $p[t]$ to fluctuate. 
To achieve stable and consistent values, the confidence is then normalized $p[t]/P_{max}$ based on the maximum $p[t]$ value of the image $P_{max}$.
Using this confidence map, we can estimate the location of the welding spot from the high confidence areas (red regions in Fig. \ref{fig:probMethod}). 
We can also predict the next location of the electric arc, labeled in green color in Fig. \ref{fig:probMethod}.

\subsubsection{Dimension Filter}\label{sec:dim_remove} 
To speed-up the computation time, regions extracted with the binarization approach that overlap with a high probability area are only considered for tracking the welding spot.
To this end, we first represent these regions by a series contours \(K = \{K_1,\ldots, K_n\} \) computed from the binary image. 
Then, the areas \(\alpha = \{\alpha_1,\ldots,\alpha_n\} \) of the contours \( K \) are computed by using the Shoelace method \( \alpha_i = \frac{1}{2}\left| {u} - \hat{u} \right| \), for $u$ and $\hat u$ defined as follows \cite{polster2006shoelace}:
\begin{equation} 
\label{eq_area}
    {u} = q_{nx} q_{iy} + \sum_{i=1}^{n-1}q_{ix} q_{(i+1)y},\quad
    \hat{u} = q_{ny} q_{ix} + \sum_{i=1}^{n-1}q_{iy} q_{(i+1)x} 
\end{equation}
where \(q = \{q_1,\ldots q_m\} \) denotes the set of $m$ image points $q_j = [q_{jx},q_{jy}]$ of the $i$th contour \(\mathnormal{K}_i\). 
We then evaluate the dimension of the areas $\alpha_i$ and apply the following condition:
\begin{equation}
{\tilde{K}}_i = 
\begin{cases}
  K_i, &\text{if $A_{min}<\alpha_i<A_{max}$}
\end{cases}
\end{equation}
where \({\tilde{K}} = \{{\tilde{K}}_1,\ldots, {\tilde{K}}_f\} \)
denote the \emph{valid} contours that can be used in the tracking algorithm, for $A_{min},A_{max}>0$ as scalars to determine this selection.

\begin{figure*}[t]
    \centering
    \includegraphics[width=1 \linewidth]{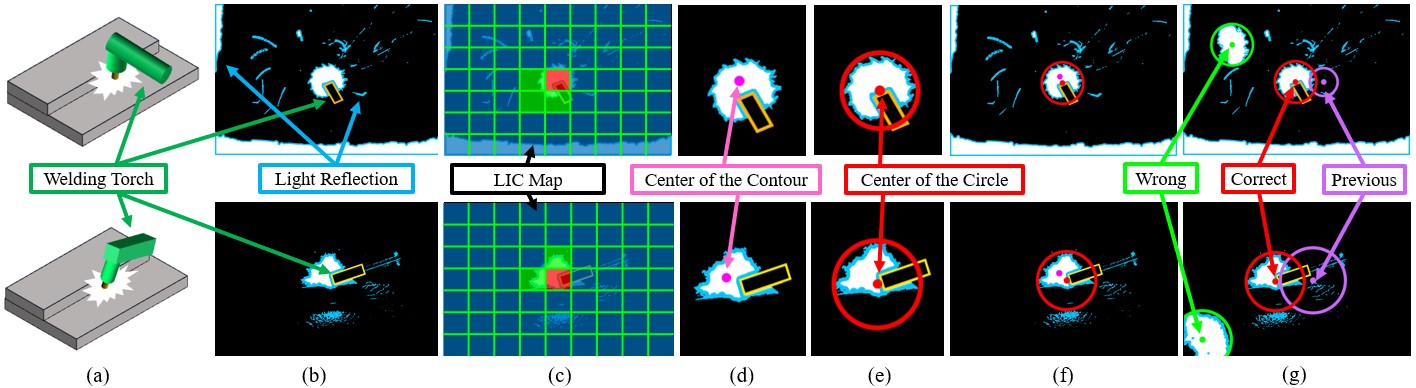}
    \caption{(a) The welding torch blocks the view; (b) Some light reflections from the surroundings share a similar light intensity level; (c) A LIC map is generated; (d) Noises in (b) are filtered out; The center of the contour labeled in pink is away from the actual center of the electrode; (e) The correct center can be found with the proposed method; (f) Comparison of the center before and after the use of the method; (g) Demonstration of the verification process.}
    \label{fig:arc_tracing}
\end{figure*}

\begin{figure}[t]
    \centering
    \includegraphics[width=1\linewidth]{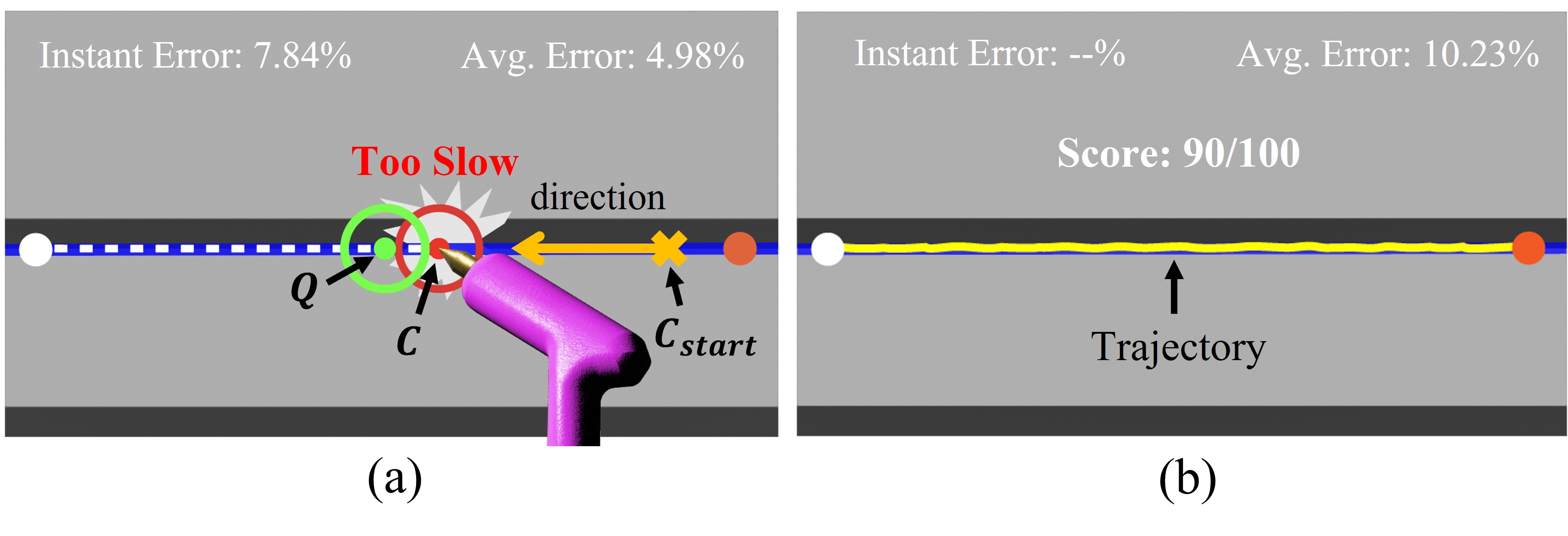}
    \caption{The conceptual interface of the XR bot trainer where $Q$ is the recommended point and $C$ the current welding spot.}
    \label{fig:interface}
\end{figure}

\subsubsection{Electric Arc Tracking}
\label{sec:elec_tracking}
During the task, the user may partially block the view of the electric arc with the torch.
This may lead to a shift between the contour's center and the actual center of the arc. 
To deal with this issue, our method uses a circle that includes all contour points \(q \) in \(\mathnormal{\alpha} \) with a minimum radius $r_i$ available. 
To this end, we model the center of the bounding circle by $C_i = [C_{ix}, C_{iy}]$.
As the region that indicates the electric arc generally produces the largest bounding circle, our algorithm verifies these regions starting from the largest circle. 
However, noises in the image may give rise to the false detection of similar-sized contours, which creates ambiguity for the system. 

Algorithm \ref{alg:circle_verify} presents the proposed verification process, which is depicted in Fig. \ref{fig:arc_tracing}.
All valid centers \(\mathnormal{C_i} \) and its radius \(\mathnormal{r_i} \) found from the contour \(\mathnormal{{\tilde{K_i}}} \) are sorted based on the size of \(\mathnormal{r_i} \) in descending order, i.e., \(\mathnormal{C}_1 \) denotes the center of the largest circle. 
By comparing the distance between the \(\mathnormal{C_i} \) with the previous state location \(\mathnormal{C_{prev}} \), the circle can be verified. 
Through iterating each \(\mathnormal{C_i} \), the largest reasonable circle around \(\mathnormal{C_{prev}} \) can be located.
Kalman filter \cite{millidge2021neural} is applied to further verify the current estimation. 
The point \(\mathnormal{C(x,y)} \) denotes the detected center of the bounding circle containing the welding spot. 
Fig. \ref{fig:interface} conceptually depicts the usage of this point in the proposed interface.

\algnewcommand{\TRUE}{\textbf{True}}
\algnewcommand{\FALSE}{\textbf{False}}
\begin{algorithm}[t]
    \caption{Electric Arc Tracking.}\label{alg:circle_verify}
    \KwIn{$C_{prev}$, ${\tilde{K}}$ \Comment{Previous center, current contours}}
    \KwOut{$C$, $r$ \Comment{Center and radius of circle}\\ }
    $N \gets \text{Compute\_Length\_of\_Array}({\tilde{K}})$\\ 
    \For{$i \gets 1$ to $N-1$}
    {
        
        $C_i$, $r_i$ = Find\_Minimum\_Circle($\tilde{K}_i$)\\
    }
    Sort\_Cicles\_by\_Maximum\_Radius($C$, $r$)\\
    \If{$C_{prev}$ is not empty}
    {
        $i \gets 1$\\
        $valid \gets False$\\
        \While{$i \le N$ \& $valid$ is $False$}
        {
            $valid$ = Compare\_Center\_Distance($C_i$, $C_{prev}$)\\
            \If{$valid$}
            {
                $C_{prev} \gets C_i$\\
                return $C_i$, $r_i$\\
            }
        }
    }
\end{algorithm}

\subsection{Motion Direction Estimation}\label{welding_dir}
The direction of the welding movement depends on the handedness of a user and the location of the seam.
For example, a right-hand welder will usually perform right to left motions in a horizontal workpiece; Vertical weld motions are typically from bottom to top.
To visually determine the direction of this welding spot, our automated system records the movement of the center \(\mathnormal{C} \) for 20 frames. 
The starting point \(\mathnormal{C_{start}} \) is computed based on sensory data captured from these first few frames. 
The moving direction is determined by comparing the signed coordinate difference between the starting point \(\mathnormal{C_{start}} \) and the current point \(\mathnormal{C} \).
For instance, if \(\mathnormal{C} \) is on the left side of \(\mathnormal{C_{start}} \) after 20-frames, it implies that the welder is going from left to right. 
This way, guidance from the XR bot trainer system can be adjusted accordingly to fit the users' torch motions.

\begin{figure*}[t]
    \centering
    \includegraphics[width=1\linewidth]{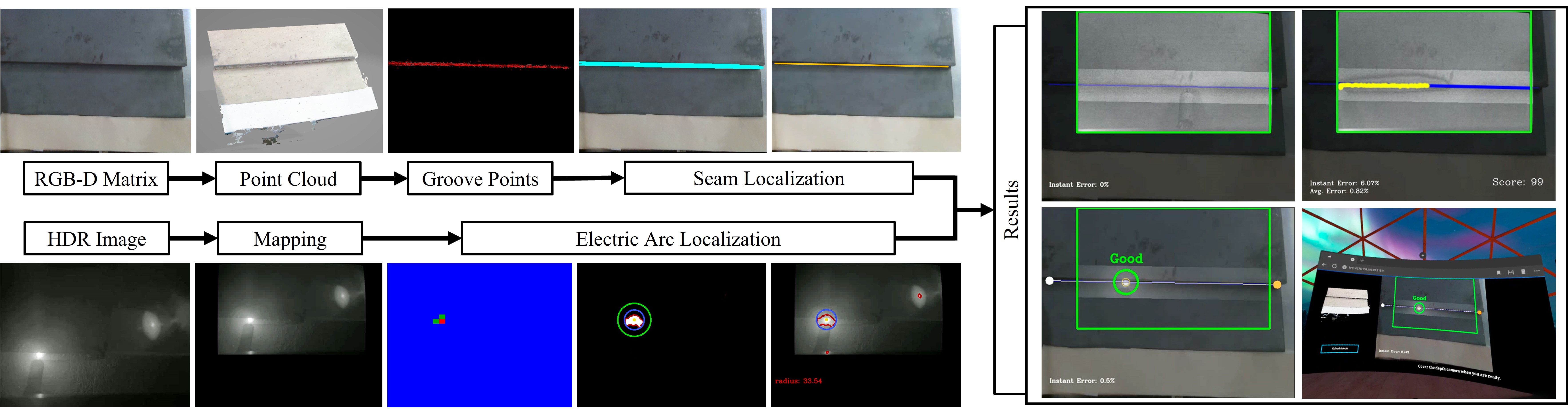}
    \caption{Overview of the welding training system. After we place a workpiece, RGB-D and HDR images are captured. The groove detection and the seam localization algorithms are executed to find the welding path. The electric arc localization is applied to the HDR image to compute the center of the arc. The yellow path indicates the welding trajectory with the seam line shown in blue. Images and results are registered into a 2D frame and uploaded to the server. Electronic devices, like a VR headset, are used to access the real-time output via a webpage.}
    \label{fig:flow}
\end{figure*}

\subsection{Virtual XR Bot Trainer}\label{sec:guidance_system}
The proposed XR bot trainer is responsible for providing instant welding guidance to the user via a head-mounted VR display.
The aim of the interface is to provide trainees with the path to be followed by the torch as well as its desired speed. 
For that, the target point \(\mathnormal{Q = [Q_x, Q_y]} \) is displayed into the VR headset as a moving circle that must be followed by the user by closely placing the electric arc's center $C$ over it.
By comparing the coordinates of the target and feedback points, we can compute the \emph{relative} motion of the torch as \(\Vec v = Q-C \), and thus, generate visual cues for the user. 
For example, if the position and velocity are close to the bot's suggestion, the point \(\mathnormal{Q} \) is displayed with a green circle; Deviations from the target location are displayed with either a red circle or a blue circle, depending on whether $C$ lags or leads the point $Q$. 

The instantaneous position error $\|Q-C\|$ and the average error \(\overline{\epsilon}\) are calculated by the XR bot trainer based on the user's motions; This information is displayed in real-time. 
After the completion of the task, the system generates a final score and plots the executed welding trajectory. 
The average error $\overline{\epsilon}$ is calculated as follows:
\begin{equation} \label{eq4}
    \overline{\epsilon} = \frac{1}{N} {\sum_{i=1}^{N} \bigg( {|C_{xi}-Q_{xi}|} / {Q_{xi}} + {|C_{yi}-Q_{yi}|} / {Q_{yi}}} \bigg)
\end{equation}
where \(N\) is the total number of target points \(Q\) involved from \(\mathnormal{C_{start}} \) to \(\mathnormal{C} \).
With this average error, the system then quantifies the task's final score as \(\gamma = 100(1-\overline{\epsilon})\).

As the proposed method is aimed at lowering the entry barrier for welding, it supports cross-platform usage. 
The real-time results and the 3D model of the workpiece are available on a webpage, which can be accessed through any electronic device via a wireless connection. 
This valuable feature enables the instructor and participants to observe the live welding process through a display device.

\section{Results}\label{sec:results}

\begin{figure}[t]
    \centering
    \includegraphics[width=1\linewidth]{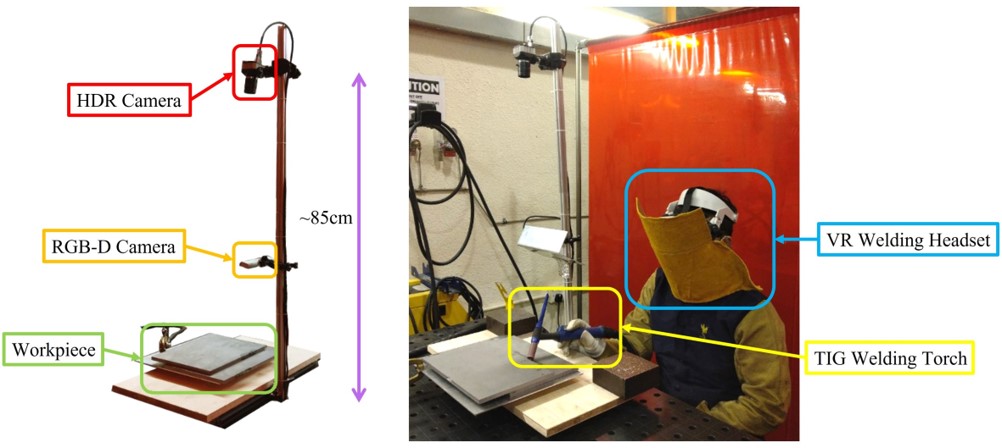}
    \caption{Experiment setup of the proposed XR welding training system.}
    \label{fig:whole_setup}
\end{figure}

\subsection{Experimental Setup}\label{sec:setup}
To validate the proposed methodology, a prototype XR interface is built and instrumented with vision sensors to capture the whole welding process.
The data processing pipeline is shown in Fig. \ref{fig:flow}, which contains an RGB-D sensor (Intel Realsense SR305), an HDR camera (New Imaging Technology MC1003), a VR headset with a controller (Oculus Quest 2), a PC (GPU RTX 3060), and a standard TIG welding torch. 
Experiments are conducted to evaluate the system, with two groups of trainees (who have no any prior welding experience) being asked to perform several tasks. 
One group learns with the traditional (i.e. current practice) approach while the other group learns with the proposed multi-sensor interface. 
In the control group, trainees practice with the instructor and receive one-to-one guidance when necessary; In the experimental group, after the instructor gives a sample demonstration, the trainees practice with the interface and receive no human guidance.
Success and failure are evaluated by the human instructor at the end of each task.
Experiments with the interface further evaluate: (1) Accuracy of the welding path, (2) accuracy of the arc location, and (3) effectiveness of the use of this system in teaching and learning.

\subsection{Experimental Design}
A horizontal fillet TIG welding is chosen for all experiments, see Fig. \ref{fig:whole_setup}.
A non-consumable electrode is used to generate the electric arc. 
The weld and the molten pool are shielded from environmental contamination by inert shielding gas \cite{weman2011welding}. 
Mild steel workpieces of $10$ mm thickness and a straight welding path are used. 
Although a fillet weld workpiece is prepared, trainees are not required to add any filler at the beginning of their training until they are capable to do so. 
To obtain a fair comparison between the two methods in terms of learning effectiveness, the task is judged as success/failure by the human instructor. 
If success is obtained before filler inclusion, then, the trainee can start adding filler in the next trial. 
If success is reached with filler inclusion, the trainee is considered to have mastered the welding skill and the practice is completed for the trainee. 

In the experiments, the system first records the traditional welding process that is conducted by the instructor. 
Then, the technique is explained to the trainees from these video recordings. 
Afterwards, trainees perform the welding task by using the interface; The instructor and other participants observe the process from a display in \emph{real-time}. 
After completion, the multimodal recordings can be replayed to analyze various details of the trainees' execution. 
As proficiency in the task is directly related to the number of repetitions a user has performed \cite{anzanello2011learning,fioretti2007organizational,newell1981mechanisms}, we quantify the system's effectiveness by computing a learning curve that relates the trial number with the obtained score.

\subsection{Performance Analysis of the Multi-Sensor Interface}\label{sec:system_analysis}

\begin{figure}[t]
    \centering
    \includegraphics[width=1\linewidth]{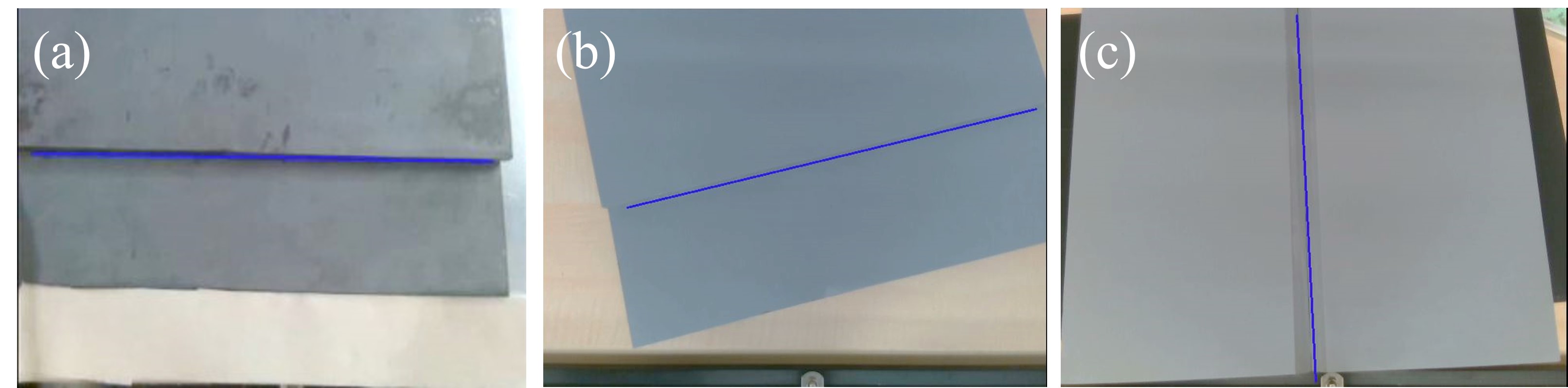}
    \caption{Various workpieces, where blue lines indicate the identified seam location. (a)-(b) Fillet workpieces; (c) Butt workpieces. 
    }
    \label{fig:seam}
\end{figure}

\begin{figure}[t]
    \centering
    \includegraphics[width=1\linewidth]{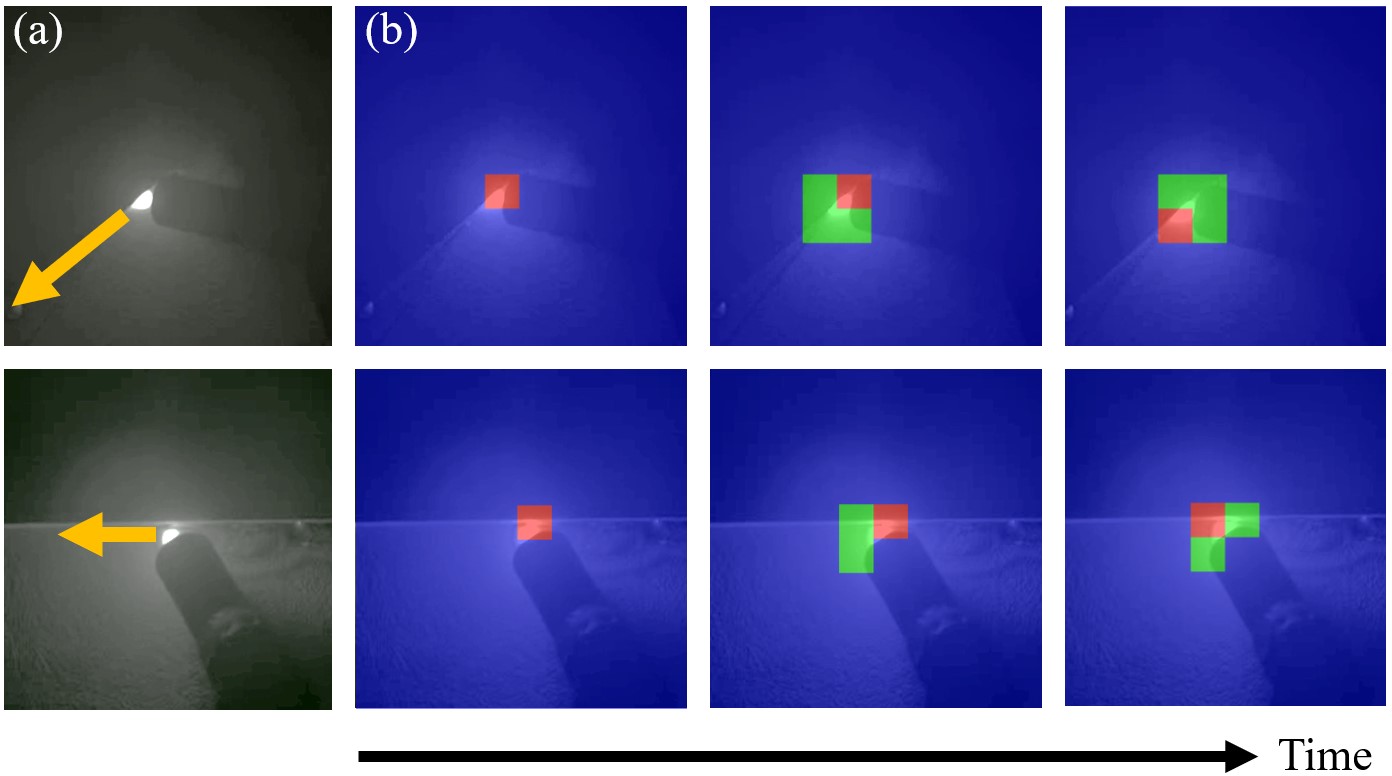}
    \caption{Visualization of the welding spot prediction that is generated by the LIC map. (a) The orange arrow indicates the welding direction; (b) Red and Green tiles indicate the potential spot location, from high to medium. Blue indicates low probability areas.}
    \label{fig:probResult}
\end{figure}

As different groove geometries may lead to different seam localization results, we conduct experiments to validate the performance of the automatic seam detection algorithm, see Fig. \ref{fig:seam}.
Fillet and butt workpieces with various orientations are used for these validation tests, which show that the algorithm can correctly locate different types of seams with various poses.
We also conducted experiments to validate the performance of the spot localization and denoising algorithms; We did it in a room with multiple \emph{unknown} light sources to test its robustness.
The LIC map is applied to the raw HDR image for noise removal, with $b=4.5$ in \eqref{eq_prob}. 
We define tiles with $p\ge 0.65$ as areas with high probability, and $p\ge0.95$ as the estimated welding spot region.
These two types of regions are depicted with green and red tiles in Fig. \ref{fig:probResult}, where we can see that the LIC map predicts the location of the welding spot from past and current intensity measurements. 

\begin{figure}[t]
    \centering
    \includegraphics[width=1\linewidth]{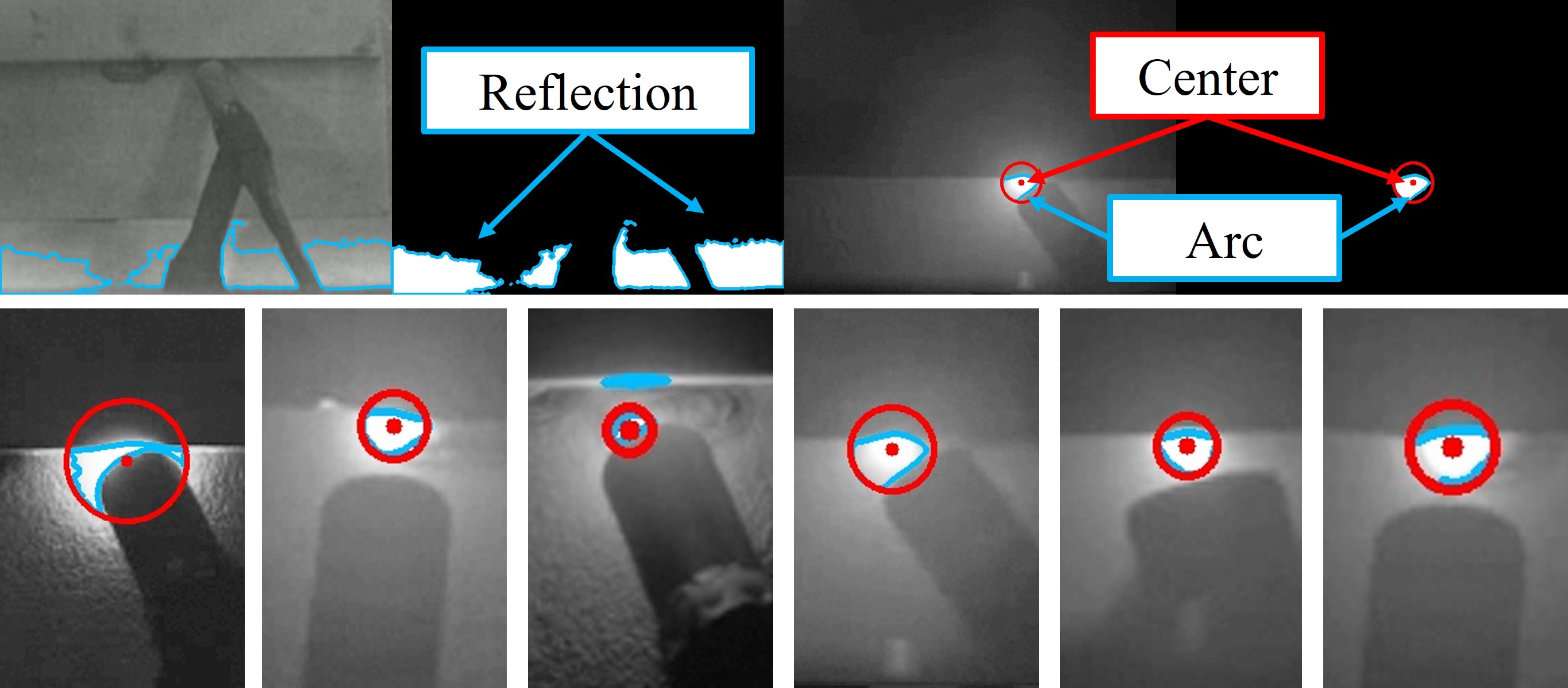}
    \caption{Visualization of the contours $K$, displayed in blue color. Red circles indicate the radius \(\mathnormal{r} \) and the center \(\mathnormal{C}\). }
    \label{fig:arc_tracing_results}
\end{figure}

\begin{figure}[t]
    \centering
    \includegraphics[width=1\linewidth]{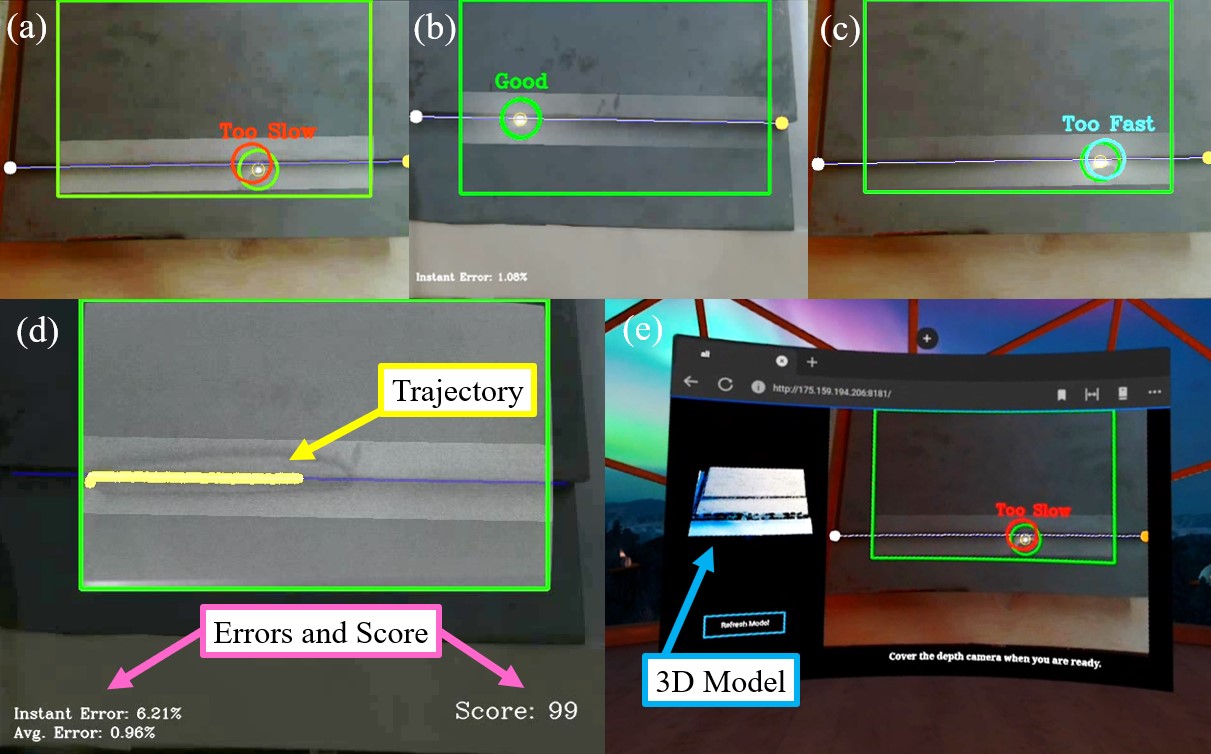}
    \caption{(a)-(c) Examples of the welding path and the guidance support provided by the XR bot trainer; The orange and white circles indicate the start and end points of the seam. (d) Welding trajectory, errors, and score of the conducted task; (e) User's view in the VR headset.}
    \label{fig:results}
\end{figure}

In our performance experiments, the welding torch is arranged at various angles and configurations, which sometimes occlude the electric arc. 
Fig. \ref{fig:arc_tracing_results} depicts configurations where the welding spot is not entirely captured by the HDR camera. 
The minimum radius circle \(\mathnormal{r} \) is formed, and thus, the center \(\mathnormal{C} \) can still be computed in these situations. 
Based on the measured trajectories of the electric arc, the bot assistant provides \emph{real-time} suggestions for user, as depicted in Fig. \ref{fig:results}. 
When the trainee has a slow pace, the suggested target point is displayed in red to alert lagging motion. 
When the speed of welding point is too high (which results in insufficient time to melt the material) the circle is displayed in blue. 
When the welding point is close to the target, the circle is shown in green. 
These features of our system provide users with valuable guidance to help them improve their technique.

\begin{figure}[t]
    \centering
    \includegraphics[width=1\linewidth]{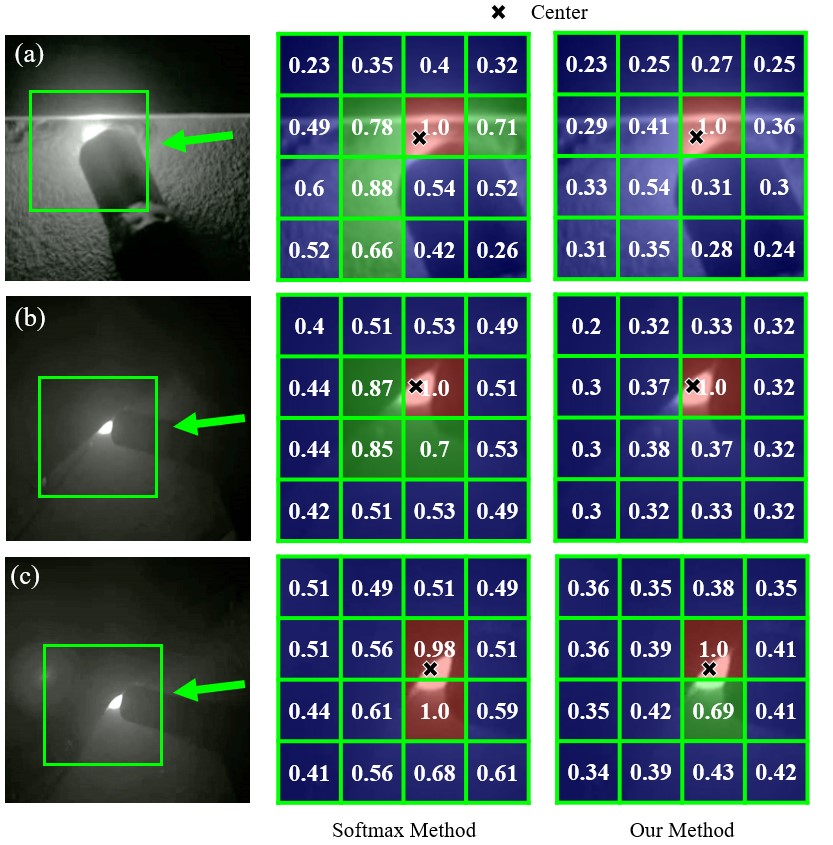}
    \caption{Comparison between the proposed LIC Map method and the traditional Softmax method in estimating the electric arc location. The black cross indicates the center of the electric arc. The green tile represents a medium-high belief ($\ge 0.65$). The red tile represents a high probability value ($\ge 0.95$). (a)--(b) The confidence value of the surrounding in the Softmax method is higher. (c) A larger error is obtained in estimating the center with the Softmax method.}
    \label{fig:probCompare}
\end{figure}

\begin{figure}[t]
    \centering
    \includegraphics[width=1\linewidth]{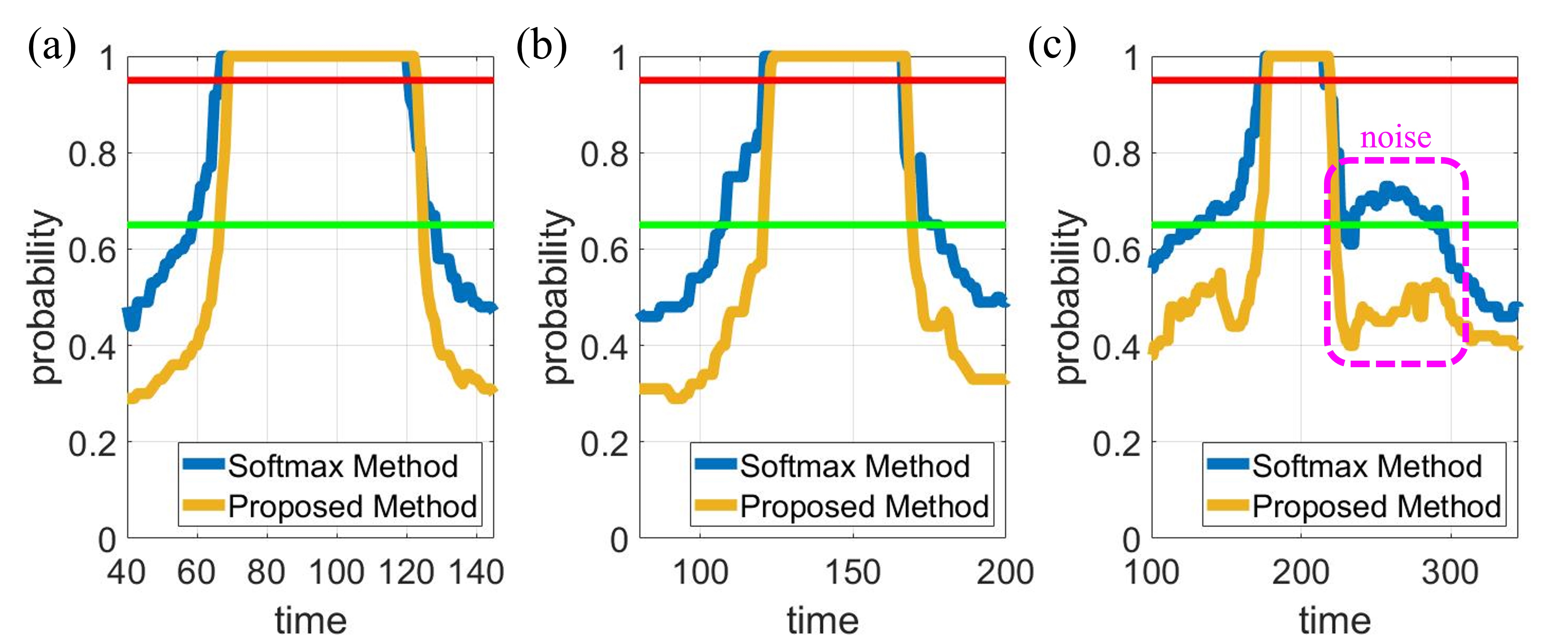}
    \caption{Evolution of the probability computed with the LIC Map method and the Softmax method for one tile. The green and red horizontal lines indicate the probability level at $0.65$ and $0.95$. The pink area is the period when noise occurs. A steeper slope and a narrower density are obtained by our proposed method. Noise is also greatly suppressed with our exponential decline function.}
    \label{fig:probCompareGraph}
\end{figure}
\begin{figure}[t]
    \centering
    \includegraphics[width=1\linewidth]{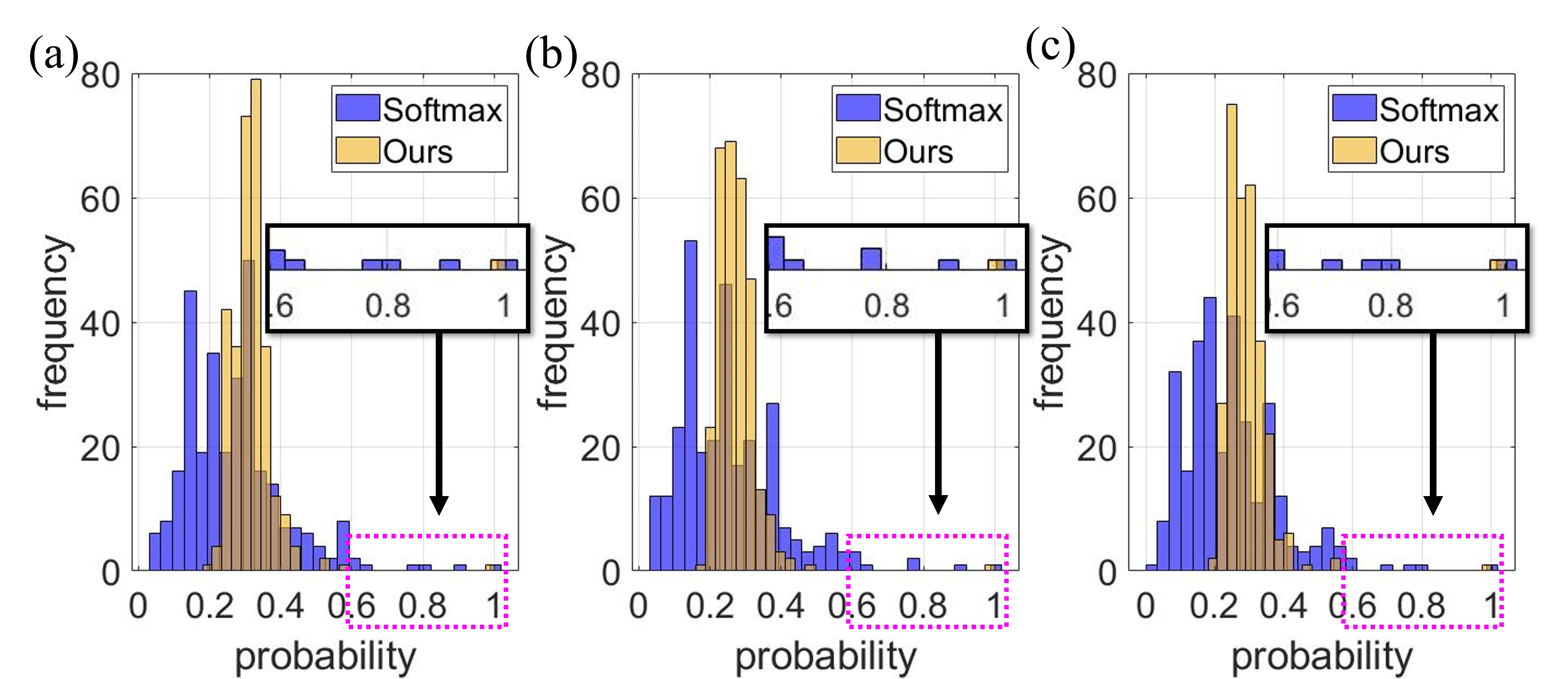}
    \caption{Comparison of the histograms obtained with the proposed LIC Map method and the Softmax method. More tiles with $\ge 0.65$ probability are found in the Softmax approach, which indicates a higher uncertainty in the estimation.}
    \label{fig:probCompareGraphHist}
\end{figure}

\begin{figure}[t]
    \centering
    \includegraphics[width=1\linewidth]{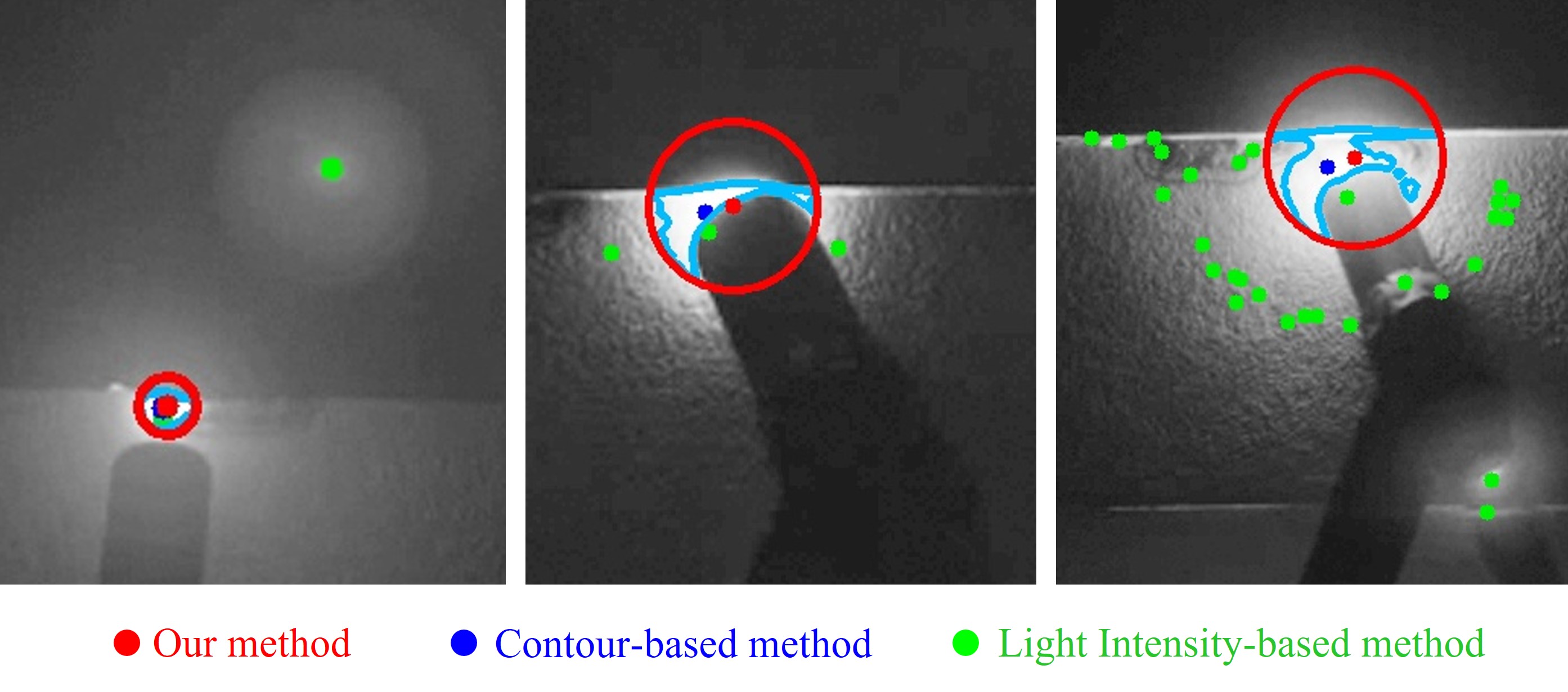}
    \caption{Comparison between the proposed method with the state-of-the-art practice in electric arc localization. The red dot is the center determined by our electric arc localization algorithm. The dark blue dot is the center found with a contour-based approach. The green dot is the center identified by the intensity-based approach.}
    \label{fig:centerCompare}
\end{figure}

\begin{figure}[t]
    \centering
    \includegraphics[width=1\linewidth]{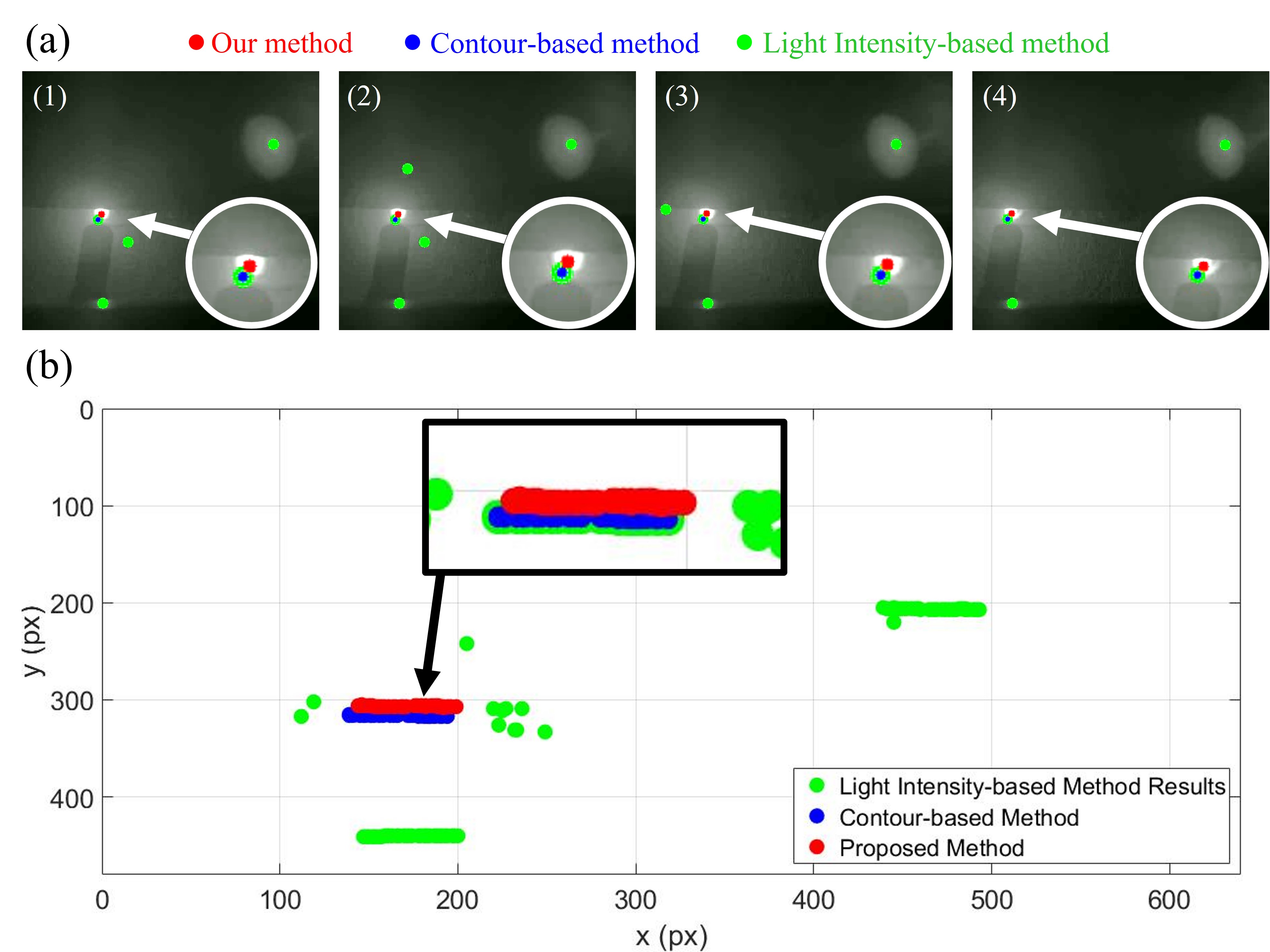}
    \caption{Quantitative comparison between the proposed method with the state-of-the-art practice in electric arc localization. (a) The welding process with results overlaid; (b) The position of the electric arc is located by three approaches. Multiple results are obtained from the light intensity-based method.}
    \label{fig:methodCompareGraph}
\end{figure}



\subsection{System Comparison}\label{sec:systemComparison}
\subsubsection{LIC Map} 
There are other prediction methods for estimating features similar to the welding spot. 
The Softmax function estimates the belief based on the current information and the weighted past events \cite{jang2016categorical, wang2018additive}. 
In contrast with this approach, our method \eqref{eq_prob} provides exponential changes in the estimated belief. 
We compare our method with the classical Softmax \cite{jang2016categorical, wang2018additive}: 
\begin{equation} \label{eq:softmax}
    p[t] = \text{norm}(\bar{\xi}) + w\cdot p[t-1]
\end{equation}
where $\text{norm}(\bar{\xi})$ is the normalized light intensity at the current time instance, and $w>0$ is a scalar weight (which we set to $w = 0.01$). 
The prediction performance of these methods can be visualized in Fig. \ref{fig:probCompare}, which shows that the tiles around the welding spot (i.e. the red tile) present a larger (undesired) belief in the Softmax approach than in our method.

To get an insight into the effect of our exponential approach, in Fig. \ref{fig:probCompareGraph} we compare the belief changes of a grid when the welding spot passes by it. 
The figure shows that the confidence level increases and decreases rapidly when the arc approaches the tile; When noises appear, our method can suppress their effect and maintain a continuous low confidence level.
The developed LIC map produces a narrow probability distribution (see Fig. \ref{fig:probCompareGraphHist}) that results in fewer potential regions for the spot localization (green and red tiles); This leads to a faster denoise process. 
With our exponential approach, we can sharpen the differences between the useful welding light area and the rest of the image. 

\subsubsection{Welding Spot Determination} 
There are various state-of-the-art approaches to locate the center of the electric arc. 
Contour-based and light intensity-based methods from the OpenCV library can be used to determine this spot. 
Fig. \ref{fig:centerCompare} compares the performance of these methods with our approach.
For fairness sake, the same input data is used for all methods, as a binary image generated by thresholding. 
In the contour-based method, the center is determined based on the largest contour's center in an image. 
In the light intensity-based approach, the center is found from all the present high light intensity regions without considering the dimension. 

The main difference between these three methods is the accuracy in determining the actual welding spot with unpredictable noise inclusion. 
In Fig. \ref{fig:centerCompare} and Fig. \ref{fig:methodCompareGraph}, the contour-based approach misidentifies the starting point of the electric arc as the welding spot. 
In the light intensity-based approach, more than one ``welding spot'' are found.
With this method, extra image processing for arc localization is normally required. 
A quantitative comparison between the three method is presented in Fig. \ref{fig:methodCompareGraph}.
A significant error can be observed in the state-of-the-art approaches.
This proves that the proposed algorithm can effective determine the the welding spot that is manipulated by the user in noisy environments.

\subsection{Effectiveness in Teaching and Learning}\label{sec:effect_analy}
\begin{figure}[t]
    \centering
    \includegraphics[width= 0.9 \linewidth]{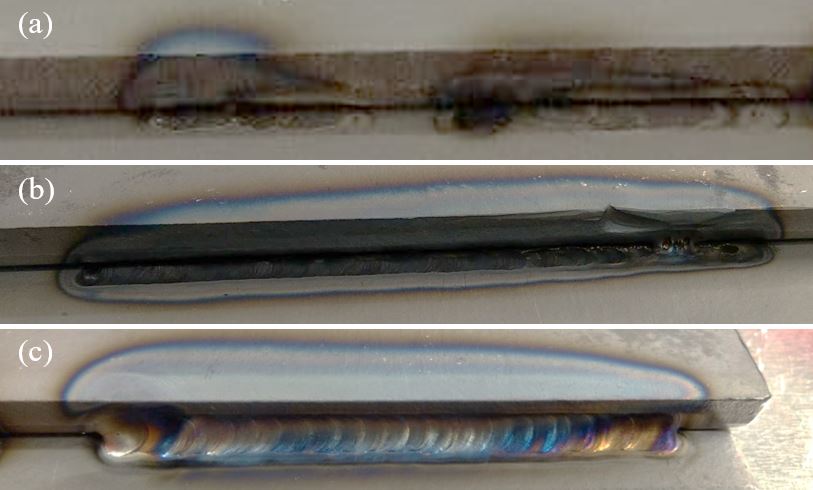}
    \caption{Examples of the trainees's performance by independently conducting welding tests. (a) Standard learning approach without filler after 5 trials; (b) Proposed method without filler after 5 trials; (c) Proposed method after 11 trials;
    (a) Standard learning approach led to consistent failures to maintain a proper movement speed; (b)--(c) Proposed method helped trainees to properly learn the technique, and thus, to master the complete TIG welding skill.}
    \label{fig:trad_new_compare}
\end{figure}

\begin{figure}[t]
    \centering
    \includegraphics[width=0.99\linewidth]{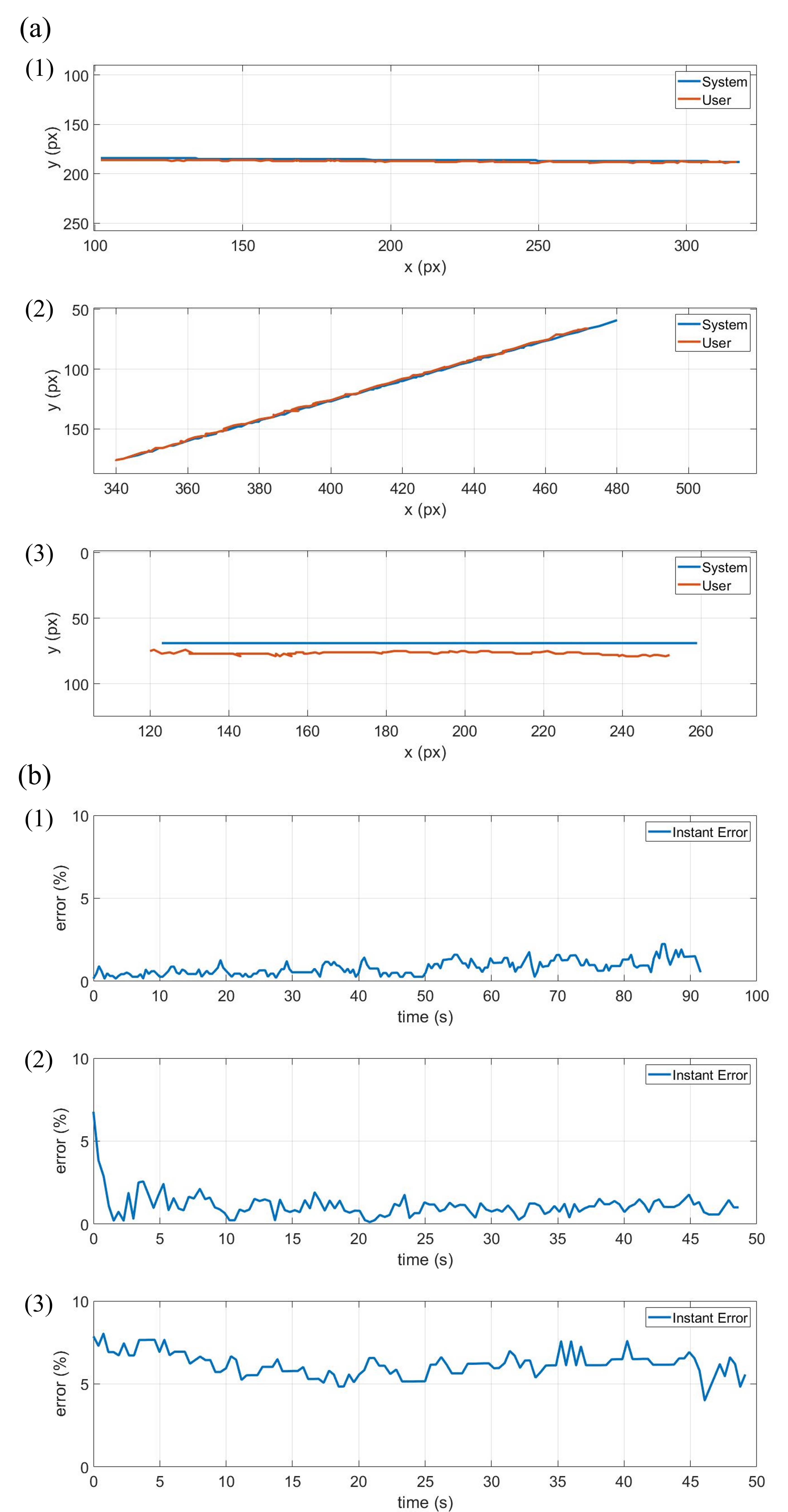}
    \caption{Sample welding trajectories performed by the trainees with the aid of our XR system. (a) Spatial x-y motions of the torch, captured during 3 different trials; (b) Instant errors of the trajectories corresponding to the motions in (a).}
    \label{fig:traj_inst_compare}
\end{figure}

As the trainees in the \emph{control} group have no prior experience with the XR bot trainer, thus, poor welding performances is observed at first. 
As they familiarize with the system, it becomes easier to perform the welding task; Filler is successfully added to the weld (a sign of skill mastery) within a short time of hands-on testing, see Fig. \ref{fig:trad_new_compare}. 
The performance of the executed welding trajectory is shown in Fig. \ref{fig:traj_inst_compare}.


\begin{figure}[t]
    \centering
    \includegraphics[width=1\linewidth]{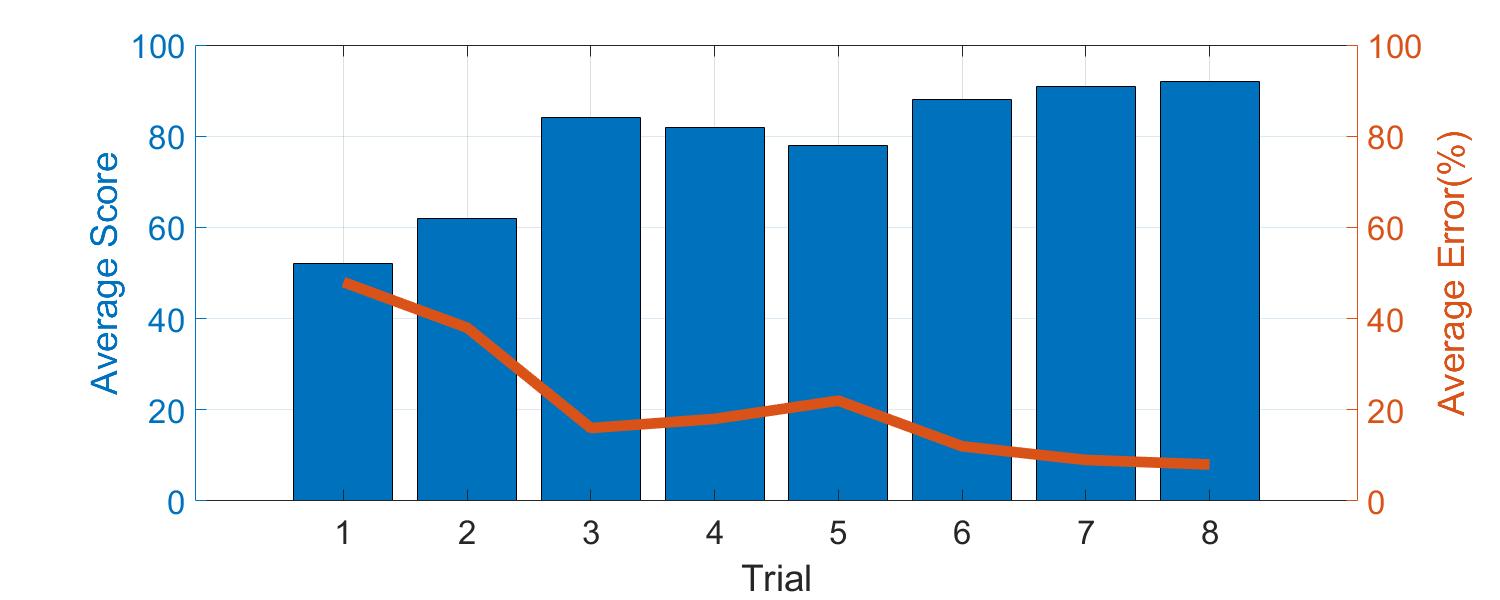}
    \caption{Average scores and errors of different welding trials conducted by the group of trainees with our system.}
    \label{fig:score_plot}
\end{figure}

Compared with the standard instructional practice, it is easier for instructors to explain the welding process and techniques to beginners with the proposed interface. 
In terms of teaching efficiency, the instructor has to keep monitoring and providing guidance to each student one by one in the traditional approach, which is a time-consuming task for them.
In the \emph{experimental} group that uses the interface, the instructor is only required to evaluate the task as success or failure once it is completed.
This greatly reduces the instructor's workload since minimal human support is needed. 
It can be concluded that our XR system can lead to a boost in teaching efficiency. 
This new approach also enables to increase the number of trainees that can participate in one session. 

With respect to the learning effectiveness, it took around 5 hours for trainees in the control group to master the welding skill while only 3 hours in the experimental group. 
On average, around 11 trials were needed for a trainee to master the TIG welding skill with our new approach; The multi-sensor interface enabled these participants to \emph{independently} conduct the training tasks.
Around 6 more trials were needed when using the traditional approach, with most participants showing a high dependency on the instructor.
The horizontal axis in Fig. \ref{fig:score_plot} indicates the number of trials conducted by the trainees in the experimental group. 
The left vertical axis represents the score obtained during the corresponding trial; The right vertical axis shows the computed average error.
As some received success at the eighth trial, no more practices are conducted.
These results demonstrate that trainees' proficiency in the task consistently improves with the number of trials. 
Effectiveness in learning grows with the use of the developed XR bot trainer.

Compared with other VR training studies, our automated XR bot assistant provides immediate \emph{real-time} support to the user, during real-world welding practice. 
The XR interface provides motion guidance and performance metrics based on sensor feedback.
Despite this instant support, our system cannot update the seam's geometric information in real-time since the welding path is computed based on the initial (static) point cloud of the workpiece.
However, the instant performance of trainees can be accessed via various types of displays that the instructor can remotely supervise. 
As virtual welding is only capable of providing a synthetic perceptual experience, trainees cannot address the typical psychological adaptation that they gain through practice in the field.

In the accompanying multimedia file, we demonstrate the performance of the system with multiple experimental videos.
\url{https://github.com/romi-lab/VR_Welding/raw/main/video.mp4}

\begin{table}[t]
\centering
\caption{Comparison with the features and functions in state-of-the-art VR methods for welding training.}
\label{table:tbl_results} 
\centering
\begin{tabular}{|M{0.1\linewidth}|M{0.2\linewidth}|M{0.12\linewidth}|M{0.2\linewidth}|M{0.12\linewidth}|}
\hline
                     & \multicolumn{2}{c|}{Virtual Welding}               & \multicolumn{2}{c|}{Real Welding}                \\ 
\hline
Methods                & Seam \& Arc Localization              & Guidance         & Seam \& Arc Localization              & Guidance  \\ 
\hline
\cite{chung2020research}                    & \checkmark             & \checkmark                   & -                 & - \\ 
\cite{isham2020mobile}                    & \checkmark             & \checkmark                   & -                 & - \\ 
\cite{yang2010virtual}                    & \checkmark             & \checkmark                   & -                 & - \\ 
\cite{kobayashi2003skill}                    & \checkmark             & \checkmark                   & -                 & - \\ 
Ours                    & -               & -                   & \checkmark              & \checkmark  \\ 

\hline
\end{tabular}
\end{table}

\begin{figure}[t]
    \centering
    \includegraphics[width=1\linewidth]{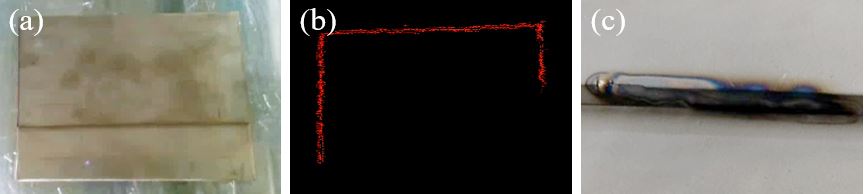}
    \caption{Some faulty examples. (a)--(b) Seam between sheet metal workpieces cannot be located as the depth difference is small; (c) Due to the limited angle of view from the system, the welding torch is hard to control with the system.}
    \label{fig:faulty}
\end{figure}

\section{Conclusion}\label{sec:conclusion}
In this paper, we presented an automated XR training bot assistant for teaching and learning arc welding tasks. 
It involves the use of a welding torch, an RGBD camera, an HDR camera, a VR headset, and image processing algorithms. 
By using this multi-sensor interface, the seam can be automatically located with 3D vision. 
The instant welding spot from the electric arc is recognized and immediate XR advice is provided to the user to improve the technique.
Task scores, errors, and paths are displayed onto the interface to provide the user with valuable feedback information. 
The effectiveness of the proposed method is experimentally validated with a group of beginners. 
Compared to the current practice, our method allows the instructor to have a clearer and more convenient way to demonstrate the process, and to quantify the trainees performance.
With the multi-sensor interface, users can independently practice the skill with minimal human supervision, hence, the proposed method has the potential to increase the number of trainees participating in a single session.

There are various limitations of the developed system.
For example, only thick materials can be considered at present. 
A thin layer such as sheet metal is too small to be recognized, as the depth difference in the groove is insignificant for the algorithm to locate the seam and it may be misjudged as one large workpiece, see e.g. Fig. \ref{fig:faulty}; The current system works for grooves of least 5-mm thickness. 
The interface may also pose difficulties in controlling the angle of the welding torch due to the limited camera view, see e.g., in Fig. \ref{fig:faulty}(c). 
This may lead to failures in joining two workpieces without a filler. 

Future work includes the integration of more sensing devices (e.g., 3D trackers) to robustly detect the pose of the torch. 
A hierarchical ranking system may be introduced to classify the users' proficiency in the task based on quantitative metrics \cite{lorenzini2022online}.
Data collected from the practice can be used to customize the bot trainer and to provide tailor-made learning exercises. 
This can help trainees to acquire the skill faster through optimizing the learning stages. 
The valuable feedback information of the interface can also be used to guide the motion of robotic welder.
A motorized robot arm is currently being developed by our team to automatically perform the task based on a simple multi-sensor system.

\bibliography{david_biblio.bib}
\bibliographystyle{IEEEtran}

\end{document}